\documentclass[12pt,a4paper,final]{iopart}

\expandafter\let\csname equation*\endcsname\relax
\expandafter\let\csname endequation*\endcsname\relax
\usepackage{amsmath}
\usepackage{amssymb}
\usepackage{graphicx}
\usepackage{dcolumn}
\usepackage{bm}
\usepackage{hyperref}
\usepackage{color}
\usepackage{xcolor}

\renewcommand\[{\begin{equation}}
\renewcommand\]{\end{equation}}

\begin{document}

\title{First-principles calculation of optical responses based on nonorthogonal localized orbitals}

\author{Chong Wang$^{1,2,3,4}$}
\author{Sibo Zhao$^{2,3}$}
\author{Xiaomi Guo$^{2,3}$}
\author{Xinguo Ren$^{5,6}$}
\author{Bing-Lin Gu$^{1,2,3}$}
\author{Yong Xu$^{2,3,4}$}
\ead{yongxu@mail.tsinghua.edu.cn}
\author{Wenhui Duan$^{1,2,3}$}
\ead{dwh@phys.tsinghua.edu.cn}

\address{$^{1}$Institute for Advanced Study, Tsinghua University, Beijing 100084, China\\
$^{2}$State Key Laboratory of Low Dimensional Quantum Physics and Department of Physics, Tsinghua University, Beijing, 100084, China\\
$^{3}$Collaborative Innovation Center of Quantum Matter, Tsinghua University, Beijing 100084, China\\
$^{4}$RIKEN Center for Emergent Matter Science (CEMS), Wako, Saitama 351-0198, Japan\\
$^{5}$Key Laboratory of Quantum Information, University of Science and Technology of China, Hefei 230026, China\\
$^{6}$Synergistic Innovation Center of Quantum Information and Quantum Physics, University of Science and Technology of China, 230026 Hefei, China
}

\date{\today}

\begin{abstract}
Based on \emph{ab initio} software packages using nonorthogonal localized orbitals, we develop a general scheme of calculating response functions. We test the performance of this method by calculating nonlinear optical responses of materials, like the shift current conductivity of monolayer WS$_2$, and achieve good agreement with previous calculations. This method bears many similarities to Wannier interpolation, which requires a challenging optimization of Wannier functions due to the conflicting requirements of orthogonality and localization. Although computationally heavier compared to Wannier interpolation, our procedure avoids the construction of Wannier functions and thus enables automated high throughput calculations of linear and nonlinear responses related to electrical, magnetic and optical material properties.
\end{abstract}

\maketitle
\section{Introduction}
Wannier functions encode all the information of a band structure for a pre-selected energy window. Moreover, due to their highly localized nature, Wannier functions can be constructed with Bloch functions at just a few reciprocal $\boldsymbol{k}$ points \cite{souza_maximally-localized_2001,marzari_maximally_1997}. After obtaining Wannier functions, information of Bloch functions at arbitrary $\boldsymbol{k}$ points can be obtained by Fourier transformation. This procedure is called Wannier interpolation \cite{marzari_maximally_2012}. Wannier interpolation captures the physics of Bloch electrons and is an efficient way to compute various physical observables of solids from first-principles calculations. Previous researches have proven its success on studying anomalous Hall effect\cite{wang_textitab_2006}, optical conductivity\cite{yates_spectral_2007}, orbital magnetization\cite{lopez_wannier-based_2012} and nonlinear optical effects including shift current\cite{wang_first-principles_2017,ibanez-azpiroz_ab_2018} and second harmonic generation\cite{wang_first-principles_2017}. 

Current dominating algorithm of obtaining Wannier functions is maximally localized Wannier function (MLWF) theory, which mixes bands in a way that maximizes the localization of Wannier functions\cite{marzari_maximally_2012}. This method is universal and serves as a post processing tool for various
\textit{ab initio} calculation packages\cite{mostofi_updated_2014}, regardless of their implementation schemes of density functional theory (DFT). However, since MLWF algorithm is essentially an optimization process, to avoid local minimum, trial Wannier functions and other parameters have to be carefully tested. This inevitable human intervention hampers its usage in high throughput materials discovery. Moreover, the optimization of Wannier functions usually breaks material symmetries that are essential to determine various material properties, including nonlinear optical responses and topological properties. One important fact is that many existing DFT packages already use localized orbitals to span the Hilbert space. These orbitals are usually optimized before the self consistent calculation and inherit the symmetry of atomic orbitals. Therefore, if the ideas of Wannier interpolation can be borrowed to use these orbitals, the disadvantages of Wannier interpolation mentioned above can be eliminated. However, localized orbitals adopted by most software packages are nonorthogonal and do not directly fit into the framework of Wannier interpolation. This practice is rationalized by two practical reasons: (i) nonorthogonal localized orbitals (NoLO) can be made more localized than orthogonal orbitals; (ii) It's generally easier to construct NoLOs, especially when a huge amount of them is needed.

In this work, we develop a general scheme for calculating response functions using NoLOs. This method bears similarity to Wannier interpolation and allows calculation of derivatives of arbitrary orders of band energies and Bloch wave functions. Therefore, it is suitable for calculation of various linear and nonlinear response functions. As an example, we calculate dielectric constant and shift current conductivity of monolayer WS$_2$, as a proof of calculations of linear response and nonlinear response functions. The obtained results are in excellent agreement with previous works. This validates our computational formalism and its implementations. Finally, we discuss the performance and symmetry properties of this method.

\section{Method}\label{sec:method}

Berry connection\cite{berry_quantal_1984} $\boldsymbol{A}_{nm}(\boldsymbol{k})=i\langle u_{n\boldsymbol{k}}|\nabla_{\boldsymbol{k}}u_{m\boldsymbol{k}}\rangle$
and its derivative $\nabla_{\boldsymbol{k}}\boldsymbol{A}$ lie at
the heart of response functions. However, due to the random phase
generated by diagonalization, direct derivative calculation by finite
difference is not possible. Wannier interpolation scheme solves this
problem by fixing the phase by choosing a definite gauge. Here we extend
this method to NoLOs. 

We label localized orbitals as $|\boldsymbol{R}n\rangle$, where $n$
is an index of the orbital inside the unit cell labeled by a lattice vector
$\boldsymbol{R}$. $n$ runs from $1$ to $N$, where $N$ is the number
of localized orbitals in one unit cell. Bloch summations of these
localized orbitals constitute a complete basis of Hamiltonian at a
$\boldsymbol{k}$ point, 
\[
|\psi_{n\boldsymbol{k}}^{(\text{L})}\rangle\equiv\sum_{\boldsymbol{R}}e^{i\boldsymbol{k}\cdot\boldsymbol{R}}|\boldsymbol{R}n\rangle,
\]
whose cell periodic part is 
\[
|u_{n\boldsymbol{k}}^{(\text{L})}\rangle\equiv e^{-i\boldsymbol{k}\cdot\hat{\boldsymbol{r}}}|\psi_{n\boldsymbol{k}}^{(\text{L})}\rangle=\sum_{\boldsymbol{R}}e^{i\boldsymbol{k}\cdot(\boldsymbol{R}-\hat{\boldsymbol{r}})}|\boldsymbol{R}n\rangle.
\]
Under this basis, the Hamiltonian matrix is
\begin{align}
\begin{split}
H_{nm}(\boldsymbol{k}) & \equiv\langle\psi_{n\boldsymbol{k}}^{(\text{L})}|\hat{H}|\psi_{m\boldsymbol{k}}^{(\text{L})}\rangle\\
 & =\sum_{\boldsymbol{R}}e^{i\boldsymbol{k}\cdot\boldsymbol{R}}\langle\boldsymbol{0}n|\hat{H}|\boldsymbol{R}m\rangle,
\end{split}\label{H-k}
\end{align}
It should be noted that the integration over the Bloch functions in the first line of Eq. (\ref{H-k}) is carried out within one unit cell, while the integration over the local orbitals in the second line is over the Born-von-Karmen supercell. Unlike the usual Wannier interpolation scheme, $|\psi_{n\boldsymbol{k}}^{(\text{L})}\rangle$
are typically neither normalized nor orthogonal to each other. An overlapping
matrix is needed to capture this property:
\begin{align}
\begin{split}
S_{nm}(\boldsymbol{k}) & \equiv\langle\psi_{n\boldsymbol{k}}^{(\text{L})}|\psi_{m\boldsymbol{k}}^{(\text{L})}\rangle\\
 & =\sum_{\boldsymbol{R}}e^{i\boldsymbol{k}\cdot\boldsymbol{R}}\langle\boldsymbol{0}n|\boldsymbol{R}m\rangle.\label{S}
\end{split}
\end{align}
Eigenstates of the Hamiltonian $\hat{H}$ are linear combinations of $|\psi_{n\boldsymbol{k}}^{(\text{L})}\rangle$,

\begin{align}
|\psi_{n\boldsymbol{k}}\rangle & =\sum_{m}V_{mn}(\boldsymbol{k})|\psi_{m\boldsymbol{k}}^{(\text{L})}\rangle,\\
\hat{H}|\psi_{n\boldsymbol{k}}\rangle & =E_{n\boldsymbol{k}}|\psi_{n\boldsymbol{k}}\rangle,
\end{align}
where $V$ and $E$ can be obtained by solving a generalized eigenvalue
problem:
\begin{equation}
H(\boldsymbol{k})v_{n\boldsymbol{k}}=E_{n\boldsymbol{k}}S(\boldsymbol{k})v_{n\boldsymbol{k}},\label{eq:gep}
\end{equation}
which is just the Schr\"odinger equation expressed in the nonorthogonal basis $|\psi_{n\boldsymbol{k}}^{(\text{L})}\rangle$. $v_{n\boldsymbol{k}}$ is a column vector and has $N$ independent solutions constituting the columns of matrix
$V$. Notice that $V$ is not a unitary matrix. Assuming the localized orbitals are linearly independent, $S$
is Hermitian and positive definite. This type of generalized
eigenvalue problem is well behaved and the eigenvector can be normalized
as 
\begin{equation}
v_{n}^{\dagger}Sv_{m}=\delta_{nm}.\label{eq:norm}
\end{equation}

$V$ can be used to express Berry connection in the basis of $|\psi_{n\boldsymbol{k}}^{(\text{L})}\rangle$,
\begin{equation}
A^{\alpha}=iV^{\dagger}S\partial_{\alpha}V+V{}^{\dagger}A^{\alpha(\text{L})}V,\label{eq:A}
\end{equation}
where 
\begin{align}
\begin{split}
A_{nm}^{\alpha(\text{L})} & \equiv i\langle u_{n\boldsymbol{k}}^{(\text{L})}|\partial_{\alpha}u_{m\boldsymbol{k}}^{(\text{L})}\rangle\\
 & =\sum_{\boldsymbol{R}}e^{i\boldsymbol{k}\cdot\boldsymbol{R}}(\langle\boldsymbol{0}n|\hat{r}^{\alpha}|\boldsymbol{R}m\rangle-R^{\alpha}\langle\boldsymbol{0}n|\boldsymbol{R}m\rangle).
\end{split}
\end{align}
where $\hat{r}$ is the position operator, $\partial_{\alpha}\equiv\partial_{k^{\alpha}}$
and $\alpha$ is the Cartesian indices. {The position matrix depends on the choice of origin, as it should, since diagonal elements of Berry connection $\boldsymbol{A}$ depend on the choice of origin. However, since only off-diagonal elements of $\boldsymbol{A}$ are needed in the expression of dielectric constant and shift current, the results are independent of the choice of origin.} For NoLOs, the position operator matrix satisfies
\begin{align}
\begin{split}
\langle\boldsymbol{0}m|\hat{r}^{\alpha}|\bar{\boldsymbol{R}}n\rangle & =(\langle\bar{\boldsymbol{R}}n|\hat{r}^{\alpha}|\boldsymbol{0}m\rangle)^{*}\\
 & =(\langle\boldsymbol{0}n|\hat{r}^{\alpha}|\boldsymbol{R}m\rangle)^{*}-R^{\alpha}\langle\boldsymbol{0}m|\bar{\boldsymbol{R}}n\rangle,
\end{split}
\end{align}
where $\bar{\boldsymbol{R}}=-\boldsymbol{R}$. 
In contrast, $\langle\boldsymbol{0}m|\hat{r}^{\alpha}|\boldsymbol{R}n\rangle$ is a Hermitian matrix (with fixed $\boldsymbol{R}$) in the usual Wannier interpolation. 

One critical issue is that the calculation of $V^{\dagger}S\partial_{\alpha}V$ still suffers
from arbitrary phases from diagonalization. Fortunately, since the
three matrices $\langle\boldsymbol{0}n|\hat{H}|\boldsymbol{R}m\rangle$,
$\langle\boldsymbol{0}n|\boldsymbol{R}m\rangle$ and $\langle\boldsymbol{0}n|\hat{r}^{\alpha}|\boldsymbol{R}m\rangle$
are known from \emph{ab initio} calculations, arbitrary derivatives
of $H$, $S$ and $A^{(\text{L})}$ are known, which makes it possible to calculate $V^{\dagger}S\partial_{\alpha}V$ \cite{van2007computation}. As an example, we calculate
$V^{\dagger}S\partial_{\alpha}V$ at $\Gamma$ ($\boldsymbol{k}=\boldsymbol{0}$) point. Differentiating
Eq. (\ref{eq:gep}), we have 
\begin{align}
\begin{split}
(\partial_\alpha H)v_n+H\partial_\alpha v_n = & (\partial_\alpha E_n)Sv_n+\\
& E_n(\partial_\alpha S)v_n+E_n S(\partial_\alpha v_n).\label{eq:dgep}
\end{split}
\end{align}
Multiplying Eq. (\ref{eq:dgep}) with $v_{n}^{\dagger}$ on the left, we have
\[
\partial_\alpha E_{n}=v_{n}^{\dagger}(\partial_\alpha H) v_{n}-E_{n}v_{n}^{\dagger}(\partial_\alpha S)v_{n}.
\]
Multiplying Eq. (\ref{eq:dgep}) with $v_{m}^{\dagger}$ ($m\ne n$) on the left, we have
\[
v_{m}^{\dagger}S\partial_\alpha v_{n}=\frac{v_{m}^{\dagger}(\partial_\alpha H)v_{n}-E_{n}v_{m}^{\dagger}(\partial_\alpha S)v_{n}}{E_n-E_m}.
\]
However, due to the intrinsic phase ambiguity of $v_{n}$, it is necessary
to introduce an extra gauge fixing condition to obtain $v_{n}^{\dagger}S \partial_\alpha v_{n}$,
\begin{equation}
v_{n\boldsymbol{0}}^{\dagger}S(\boldsymbol{0})v_{n\boldsymbol{k}}\in\mathbb{R}.\label{eq:phase}
\end{equation}
Since at $\Gamma$ point $v_{n\boldsymbol{0}}^{\dagger}S(\boldsymbol{0})v_{n\boldsymbol{0}}=1$, $v_{n\boldsymbol{0}}^{\dagger}S(\boldsymbol{0})v_{n\boldsymbol{k}}$ is actually positive around $\Gamma$. Differentiating both Eq. (\ref{eq:norm}) and Eq. (\ref{eq:phase}),
we have 
\[
v_{n}^{\dagger}S\partial_\alpha v_n=-\frac{1}{2}v_{n}^{\dagger}(\partial_\alpha S)v_n.
\]
We have assumed in the above equations that $n$ is not degenerate. Degenerate eigenvalues can be treated in principle\cite{andrew1998computation} but are not important in our calculations\cite{wang_first-principles_2017}. In the calculations, the procedure outlined above is done independently at every $\boldsymbol{k}$ point, owing to the property that response functions can be expressed in a way that a global smooth gauge is not needed\cite{wang_first-principles_2017}.

Every ingredient needed for linear response are obtained at this point. However, we still need $\nabla_{\boldsymbol{k}}\boldsymbol{A}$ for nonlinear responses. Differentiating Eq. (\ref{eq:A}) again, we have
\[
\begin{split}
\partial_{\beta}A^{\alpha}  = &i(\partial_{\beta}V^{\dagger})S(\partial_{\alpha}V)+iV^{\dagger}(\partial_{\beta}S)(\partial_{\alpha}V)\\
&+iV^{\dagger}S\partial_{\beta}\partial_{\alpha}V+(\partial_{\beta})V^{\dagger}A^{\alpha(\text{L})}V\\
&+V^{\dagger}(\partial_{\beta}A^{\alpha(\text{L})})V+V^{\dagger}A^{\alpha(\text{L})}(\partial_{\beta}V).
\end{split}
\]
The corresponding result for orthonormal Wannier functions\cite{wang_first-principles_2017} is reproduced by choosing $S = I$. The calculation of $V^{\dagger}S\partial_{\beta}\partial_{\alpha}V$ follows
exactly the same logic of calculating $V^{\dagger}S\partial_{\alpha}V$.
However, it is not difficult to imagine how complex the final expression
would become. Therefore, we introduce in the appendix an iteration
procedure to calculate derivatives of $v_n$ of arbitrary orders for nondegenerate
$E_{n}$.

Another way to find Berry connection is to do orthogonalization by $S^{-1/2}$ in Eq. (\ref{eq:gep}) and use the usual Wannier interpolation developed previously, but derivatives of $S^{-1/2}$ will be needed to calculate derivatives of the transformed Hamiltonian. This orthogonalization scheme will be developed and discussed in detail in Sec. \ref{sec:orthogonalization}.

Finally, we remark that since the full spectrum is needed at every $\boldsymbol{k}$ point, this method includes a non-self-consistent calculation at every $\boldsymbol{k}$ point, in contrast to the usual Wannier interpolation scheme. Henceforth, we will refer to the scheme developed here as NoLO-based method and the usual Wannier interpolation as MLWF-based interpolation.

\section{Shift Current of Monolayer WS$_2$}

\subsection{Background and Computation Details}

Shift current\cite{von_baltz_theory_1981,sipe_second-order_2000,young_first_2012,tan_shift_2016} is a second-order bulk photovoltaic effect arising from the difference of real space positions of Bloch electrons between valence band and conduction band,
\begin{equation}
J^{\alpha}=\sigma^{\alpha\beta\beta}(\omega)E^{\beta}(\omega)E^{\beta}(-\omega),\label{eq:shift_current-def}
\end{equation}
where shift current conductivity $\sigma^{\alpha\beta\beta}$ is given by\cite{von_baltz_theory_1981,sipe_second-order_2000,young_first_2012,tan_shift_2016}
\begin{equation}
\sigma^{\alpha\beta\beta}(\omega)=\frac{2g_s\pi e^3}{\hbar^2}\int\frac{d^3\boldsymbol{k}}{(2\pi)^3}\sum_{n,m}f_{nm}I_{nm}^{\alpha\beta\beta}\delta(\omega_{nm}-\omega),\label{eq:shift-current}
\end{equation}
where $g_{s}$ is the spin degeneracy, $\hbar\omega_{nm}=E_{n}-E_{m}$ represents photon energy,  $f_{nm} = f(E_{n}) - f(E_{m})$ and $f$ is Fermi-Dirac distribution. The integrand $I_{nm}^{\alpha\beta\beta}$ is composed of transition rate from band $m$ to band $n$ and shift vector between the two bands. $I_{nm}^{\alpha\beta\beta}$ can be written out with Berry connections and derivatives of Berry connections:
\begin{equation}
I_{nm}^{\alpha\beta\beta}=\text{Im}[A_{mn}^{\beta}A_{nm;\alpha}^{\beta}],
\end{equation}
where $A_{nm;\alpha}^{\beta}=\partial_\alpha A_{nm}^{\beta}-i(A_{nn}^{\alpha}-A_{mm}^{\alpha})A_{nm}^{\beta}$.
These quantities can be calculated using the method described in Sec.\ref{sec:method}. Then an numerical integration would produce the result of $\sigma^{\alpha\beta\beta}$. Since a $\delta$ function is present in the expression of $\sigma^{\alpha\beta\beta}$, a very fine $\boldsymbol{k}$ mesh is needed to achieve convergence.

Since monolayer WS$_2$ has the point group $D_{6h}$, there's only one independent shift current conductivity component $\sigma^{yyy}$\cite{bilbao,wang_first-principles_2017}\footnote{We ignore any shift current conductivity components containing direction $z$ due to the 2D nature of monolayer WS$_2$.}. Following the convention of Ref. \cite{wang_first-principles_2017}, we choose a two dimensional (2D) definition of current. {Two DFT packages are utilized in the calculations: the full-potential, all-electron \textsc{fhi-aims} package\cite{blum_ab_2009} and the pseudopotential-based \textsc{openMX}\cite{ozaki_variationally_2003,ozaki_numerical_2004} package. For comparison, MLWF-based calculations are also carried out with \textsc{vasp}\cite{kresse1993ab,kresse_efficiency_1996} and \textsc{wannier90}\cite{mostofi_updated_2014} packages.} In all the calculations, slab model is used to characterize monolayer WS$_2$ with a vacuum layer thicker than 15\AA. A $\boldsymbol{k}$ grid $12\times12\times1$ is used to sample the Brillouin Zone in self consistent calculations and a much finer $\boldsymbol{k}$ grid $400\times400\times1$ is used to perform the numerical integration in the expression of shift current conductivity. Interaction effects are captured in Perdew-Burke-Ernzerhof (PBE) exchange-correlation functional\cite{perdew_generalized_1996}. To expand the Kohn-Sham wave functions, we choose the so-called ``tight" numerical settings in \textsc{fhi-aims} calculations, pseudo atomic basis ``W7.0-s3p2d2f1'' and ``S7.0-s3p3d2f1'' in \textsc{openMX} calculations and plane wave basis with an energy cut of 258.689 eV in \textsc{vasp} calculations. Spin-orbit interaction is not included and {$\delta$ function is simulated using the following numerical approximation 
\[
\delta(x)=\lim_{\epsilon\to0}\frac{1}{\epsilon\sqrt{\pi}}e^{-x^2/\epsilon^2},
\]
where the broadening factor $\epsilon$ is chosen to be $0.1$ eV.}

\subsection{Results}

\begin{figure}
\includegraphics[width=0.7\columnwidth]{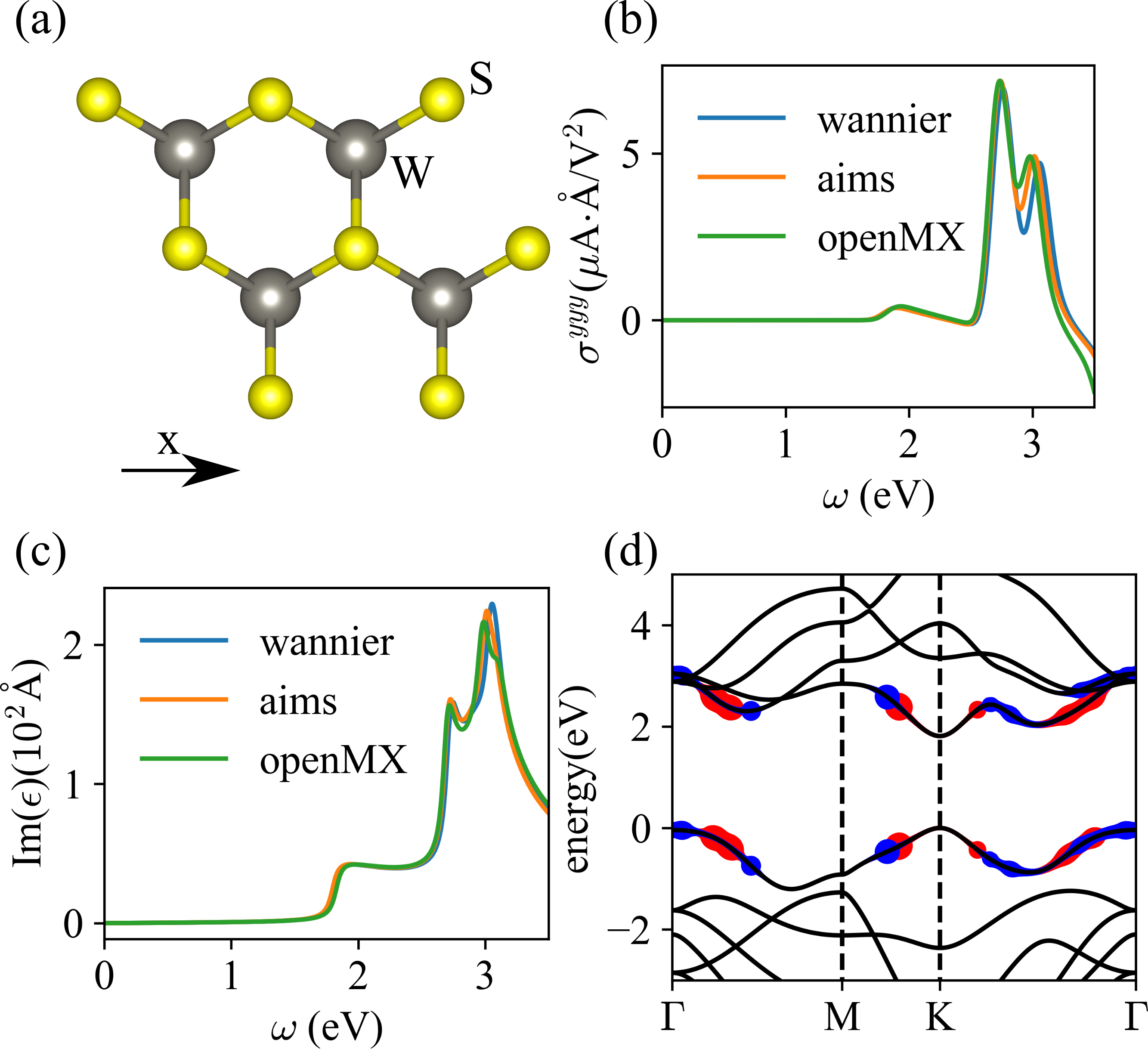}
\centering
\caption{(Color online) Shift current conductivity of monolayer WS$_2$. (a) Top view of monolayer WS$_2$. (b) $\sigma^{yyy}$ of monolayer WS$_2$ as a function of photon energy calculated with MLWF-based interpolation and NoLO-based method. {Two software packages (\textsc{fhi-aims} and \textsc{openMX}) are used in NoLO-based method.} (c) Imaginary part of dielectric constant of monolayer WS$_2$ as a function of photon energy calculated with MLWF-based interpolation and NoLO-based method. Contributions to the first and second peaks are decomposed to bands, shown by size of red and blue dots, respectively, in (d). We have adopted 2D versions of current and polarization, thus an extra \AA~is introduced in the units of $\sigma$ and $\epsilon$. (d) Band structure of WS$_2$.\label{fig:WS2}}
\end{figure}

The shift current conductivity of monolayer WS$_2$ is presented in Fig. \ref{fig:WS2}(a), compared with MLWF-based method introduced in Ref. \cite{wang_first-principles_2017} with the same parameters. Despite using different software packages and DFT schemes, the calculated shift current conductivity curves are almost the same. Optical absorptions, represented by imaginary part of dielectric constant calculated by both MLMF-based interpolation and the method developed here, are plotted in Fig. \ref{fig:WS2}(b). Both dielectric constant and shift current conductivity have two peaks at 2.75eV and 3.05eV respectively. While for dielectric constant, the peak at 3.05eV is higher than that at 2.75eV, an opposite feature is observed for shift current conductivity. This difference should be attributed to the difference of shift vectors. The contribution of these two peaks to dielectric constant can be decomposed to individual bands and is shown in Fig. \ref{fig:WS2}(c). Sizes of red and blue dots represent contributions to the 2.75eV peak and 3.05eV absorption peak respectively. It is obvious that both peaks are mainly contributed by the highest valence band and lowest
four conduction bands around $\Gamma$ and $K$ points.

\subsection{Discussion of the Method}

\begin{figure}
\includegraphics[width=0.7\columnwidth]{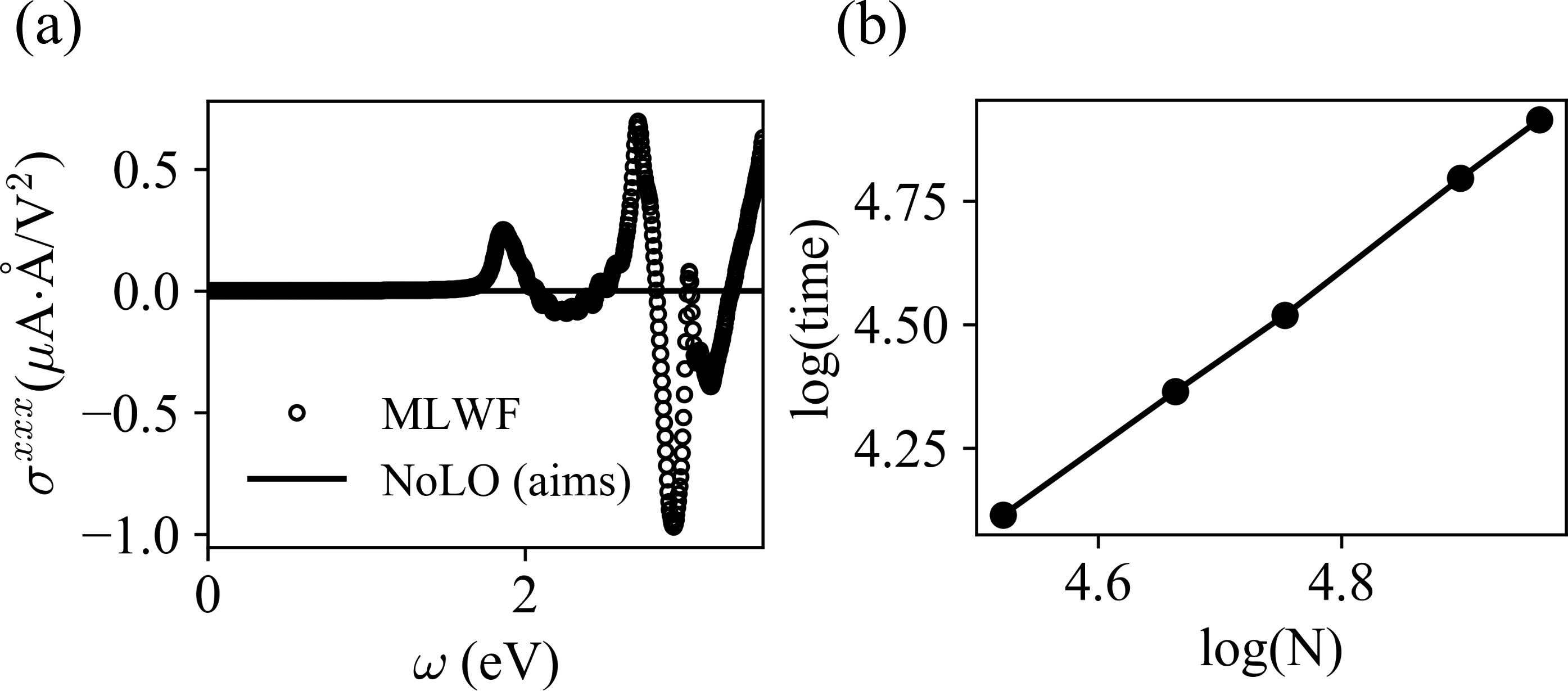}
\centering
\caption{(a) Calculation of symmetry forbidden components of shift current conductivity of monolayer WS$_2$. (b) The computational time with respect to number of NoLOs ($N$). Black dots are actual data. Slope of the line is roughly 2.\label{fig:performance}}
\end{figure}

MLWFs are known to break symmetry slightly, which is revealed by small avoid crossings in the interpolated band structure where they should have been direct crossings. This behavior results in a small but nonvanishing value for symmetry forbidden components of shift current conductivity even for well convergent MLWFs. This problem does not arise for NoLO-based method since symmetry is enforced in \emph{ab initio} calculations by only sampling the irreducible Brillouin Zone in the calculation. Fig. \ref{fig:performance}(a) shows a forbidden component of shift conducitivity of WS$_2$ ($\sigma^{xxx}$) calculated by the current method and MLWF-based interpolation. It can be observed that while MLWF-based interpolation gives a value around 1$\mu$A$\cdot$\AA/V$^2$ for this component, results of NoLO-based method are vanishingly small (no more than $10^{-4}\mu$A$\cdot$\AA/V$^2$). Therefore, NoLO-based method preserves symmetry properties quite well.

Compared to MLWF-based interpolation, NoLO-based calculation is computationally heavier, since \emph{ab initio} packages, especially all-electron full-potential packages, need to use many NoLOs to span the Hilbert space, while MLWFs are usually only constructed for bands near the Fermi surface. Therefore, it is necessary to test the scaling behavior of this method with respect to the number of NoLOs. This scaling behavior is presented in Fig. \ref{fig:performance}(b). It is observed that computation time roughly scales as $\text{O}(N^2)$, where $N$ is the number of NoLOs. This is quite unexpected since diagonalization scales as $\text{O}(N^3)$. A closer analysis of the performance reveals that the computation is dominated by Fourier transformations in calculating $H$, $S$, $A^{(\text{L})}$ and their derivatives. Therefore, for common bulk materials, we can safely assume the time complexity is $\text{O}(N^2)$. {There is another source of performance difference between the MLWF-based method and the NoLO-based method. Due to the orthogonality of MLWFs, the Fourier transformation of $S$ is trivial in the MLWF-based method and thus saves one third of the time compared to NoLO-based method, even if they use the same number of orbitals.}

\section{Alternative scheme by orthogonalization\label{sec:orthogonalization}}

Although $|\psi^{(\text{L})}_{n\boldsymbol{k}}\rangle$ are not orthogonal to each other by definition, an orthogonalization can bring it to an orthogonal basis
\begin{equation}
    |\psi^{(\text{O})}_{n\boldsymbol{k}}\rangle=
    \sum_m |\psi^{(\text{L})}_{m\boldsymbol{k}}\rangle S^{-1/2}_{mn}(\boldsymbol{k}).
\end{equation}
The Hamiltonian in the basis of $|\psi^{(\text{O})}_{n\boldsymbol{k}}\rangle$ is therefore related to $H^{(\text{L})}(\boldsymbol{k})$ as
\begin{equation}
    H^{(\text{O})}(\boldsymbol{k})=S^{1/2}(\boldsymbol{k})H^{(\text{L})}(\boldsymbol{k})S^{-1/2}(\boldsymbol{k}).\label{HO-HW}
\end{equation}
Notice that the principle square root of $S(\boldsymbol{k})$ is well defined since $S(\boldsymbol{k})$ is positive definite. In this way, MWLF-based interpolation method can be directly applied to calculate nonlinear optical responses.

However, one key ingredient of MWLF-based interpolation is the derivative of $H^{(\text{O})}(\boldsymbol{k})$ with respect to $\boldsymbol{k}$. By Eq. (\ref{HO-HW}), this issue reduces to the calculation of derivatives of arbitrary order of $S^{-1/2}(\boldsymbol{k})$ with respect to $\boldsymbol{k}$. This can be done as follows. By differentiating $S^{-1/2}S^{-1/2}=S^{-1}$, we obtain
\begin{equation}
    (\partial_{\alpha}S^{-1/2})S^{-1/2}+S^{-1/2}\partial_{\alpha}S^{-1/2}=\partial_{\alpha}S^{-1}.\label{dS1/2-1}
\end{equation}
$\partial_{\alpha}S^{-1}$, on the other hand, can be calculated by differentiating $SS^{-1}=I$,
\begin{align}
    (\partial_{\alpha}S)S^{-1}+S\partial_{\alpha}S^{-1}=0
\end{align}
and getting
\begin{align}
    \partial_{\alpha}S^{-1} = -S^{-1}(\partial_{\alpha}S)S^{-1}.
\end{align}

Eq. (\ref{dS1/2-1}) is in the form of a Sylvester equation\cite{sylvester}. Sylvester equation, with positive definite $S$ here, has one and only one solution. This solution, accessible from existing code\cite{laug}, is the desired first order derivative of $S^{-1/2}(\boldsymbol{k})$. By differentiating $S^{-1/2}S^{-1/2}=S^{-1}$ and $SS^{-1}=I$ to second order, the second order derivative of $S^{-1/2}(\boldsymbol{k})$ can be obtained in a similar manner. In a similar way, arbitrary derivative of $S^{-1/2}(\boldsymbol{k})$ can be obtained through a recursive calculation.

Orthogonalization method presented in this section has the advantage of being directly connected to previous Wannier interpolation methods. However, the pervasive usage of matrix inversion makes this method slightly slower and less numerical stable than the method presented in Sec. \ref{sec:method}. 

\section{Conclusion}

Real space localized orbital-based \emph{ab initio} packages have the potential of achieving $\text{O}(N)$ computational resource scaling with respect to number of atoms. In addition, vacuum can be treated without extra computational cost in these packages, making them competitive tools in research for low dimensional materials. Here we demonstrate these packages can be more powerful by extending Wannier interpolation to NoLOs.

Although computationally heavier compared to MLWF-based interpolation, NoLO-based scheme developed in this work avoids human intervention and can be used in high throughput material discovery. The correctness of this scheme is proved by calculating shift current conductivity of monolayer WS$_2$. This NoLO-based method is quite general and can be used to calculate different kinds of linear responses and nonlinear responses in different research fields\cite{tokura_nonreciprocal_2018}, including anomalous Hall effect (see also Ref. \cite{lee_tight-binding_2018}\footnote{{Ref. \cite{lee_tight-binding_2018} takes a linear response approach based on Kubo formula. The central objects in Ref. \cite{lee_tight-binding_2018} is the optical matrix.}}) for a optical matrix based approach), nonlinear Hall effect, orbital magnetization and second harmonic generation.

\section*{Acknowledgements}

We thank William P. Huhn, Volker Blum and Honghui Shang for helpful discussions on interfacing our code with \textsc{fhi-aims}. We are also grateful for Chi-Cheng Lee and T. Ozaki for their help in interfacing our code with \textsc{openMX}. Our code is written in julia programming language\cite{julia}. C.W., S.Z., X.G., B.-L.G., Y.X. and W.D. acknowledge support from the Ministry of Science and Technology of China (Grants No. 2016YFA0301001, No. 2018YFA0307100, No. 2018YFA0305603 and No. 2017YFB0701502), the National Natural Science Foundation of China (Grants No. 11674188, No. 11334006, No. 11874035, and No. 51788104) and the Beijing Advanced Innovation Center for Future Chip (ICFC). X.R. acknowledges support from Chinese National Science Foundation (Grant No. 11574283).

\appendix
\section{Derivative of Eigenvalues and Eigenvectors of Generalized Eigenvalue Problem}
In this appendix, we derive derivatives of eigenvectors and eigenvalues of arbitrary orders of the generalized eigenvalue problem

\begin{equation}
Hx=\lambda Sx,\label{app:eq:gep}
\end{equation}

To derive higher order derivatives of eigenvectors, we follow the
same pattern as in the main text. However, the expressions become
extremely long after differentiating, thus we introduce some useful
symbols here. To motivate the symbols, we first try to differentiate
$A_{1}(k)A_{2}(k)A_{3}(k)$ to second order. The result is $A_{1}^{(2)}(k)A_{2}(k)A_{3}(k)+A_{1}(k)A_{2}^{(2)}(k)A_{3}(k)+A_{1}(k)A_{2}(k)A_{3}^{(2)}(k)+2A_{1}^{(1)}(k)A_{2}^{(1)}(k)A_{3}(k)+2A_{1}^{(1)}(k)A_{2}(k)A_{3}^{(1)}(k)+2A_{1}(k)A_{2}^{(1)}(k)A_{3}^{(1)}(k)$,
which can be written concisely as $\sum_{P_{3}^{2}}C_{P_{3}^{2}}A_{1}^{(P_{3}^{2}(1))}A_{2}^{(P_{3}^{2}(2))}A_{3}^{(P_{3}^{2}(3))}$
with 
\[
P_{3}^{2}\in\{(2,0,0),(0,2,0),(1,1,0),(1,0,1),(0,1,1),(0,0,2)\}
\]
 and 
\[
C_{P_{3}^{2}}=\{1,1,2,2,2,1\}
\]
 respectively. These symbols can be readily extended for multivariable
differentiation. For example, differetiating $A_{1}(k_{1},k_{2})A_{2}(k_{1},k_{2})A_{3}(k_{1},k_{2})$
with respect to $k_{1}$ and $k_{2}$, we have $\sum_{P_{3}^{(1,1)}}C_{P_{3}}A_{1}^{(P_{3}^{(1,1)}(1))}A_{2}^{(P_{3}^{(1,1)}(2))}A_{3}^{(P_{3}^{(1,1)}(3))}$,
where 
\begin{align*}
P_{3}^{(1,1)} & \in\{((1,1),(0,0),(0,0)),((1,0),(0,1),(0,0))((1,0),(0,0),(0,1)),\\
 & ((0,1),(1,0),(0,0)),((0,0),(1,1),(0,0))((0,0),(1,0),(0,1)),\\
 & ((0,1),(0,0),(1,0)),((0,0),(0,1),(1,0))((0,0),(0,0),(1,1))\}
\end{align*}
and 
\[
C_{P_{3}^{(1,1)}}=1
\]
for all $P_{3}^{(1,1)}$. General expressions of $P$ and $C_{P}$
can be derived using binomial polynomial. $\{P_{N}\}$ can be ordered such that a specific
combination can be referenced. We do not try to order the set generally
but define $P_{N}^{1}$ to be all the differentiation going to $A_{1}$,
and $P_{N}^{-1}$ to be all differentiation going to $A_{N}$.
It is obvious $C_{P_{N}^{1}}=C_{P_{N}^{-1}}=1$.

Now we are ready to compute the derivatives.

Differentiate Eq. (\ref{app:eq:gep}), we have
\[
\sum_{P_{2}}C_{P}H^{(P)}x_{i}^{(P)}=\sum_{P_{3}}C_{P}\lambda_{i}^{(P)}S^{(P)}x_{i}^{(P)},
\]
 which generates the following two equation
\begin{align}
\sum_{P_{2}}C_{P}x_{i}^{\dagger}H^{(P)}x_{i}^{(P)} & =\sum_{P_{3}}C_{P}\lambda_{i}^{(P)}x_{i}^{\dagger}S^{(P)}x_{i}^{(P)},\\
\sum_{P_{2}}C_{P}x_{j}^{\dagger}H^{(P)}x_{i}^{(P)} & =\sum_{P_{3}}C_{P}\lambda_{i}^{(P)}x_{j}^{\dagger}S^{(P)}x_{i}^{(P)}.
\end{align}
Eliminating zeros, we have 
\begin{align}
\sum_{P_{2}\ne P_{2}^{-1}}C_{P}x_{i}^{\dagger}H^{(P)}x_{i}^{(P)} & =\sum_{P_{3}\ne P_{3}^{-1},P_{3}^{1}}C_{P}\lambda_{i}^{(P)}x_{i}^{\dagger}S^{(P)}x_{i}^{(P)}+\lambda_{i}^{(P_{3}^{1}(1))},\\
\sum_{P_{2}\ne P_{2}^{-1}}C_{P}x_{j}^{\dagger}H^{(P)}x_{i}^{(P)}+\lambda_{j}x_{j}^{\dagger}Sx_{i}^{(P_{2}^{-1}(2))} & =\sum_{P_{3}\ne P_{3}^{-1},P_{3}^{1}}C_{P}\lambda_{i}^{(P)}x_{j}^{\dagger}S^{(P)}x_{i}^{(P)}+\lambda_{i}x_{j}^{\dagger}Sx_{i}^{(P_{3}^{-1}(3))}.
\end{align}
Thus, 
\begin{align}
\lambda_{i}^{(P_{3}^{1}(1))} & =\sum_{P_{2}\ne P_{2}^{-1}}C_{P}x_{i}^{\dagger}H^{(P)}x_{i}^{(P)}-\sum_{P_{3}\ne P_{3}^{-1},P_{3}^{1}}C_{P}\lambda_{i}^{(P)}x_{i}^{\dagger}S^{(P)}x_{i}^{(P)},\\
x_{j}^{\dagger}Sx_{i}^{(P_{3}^{-1}(3))} & =(\sum_{P_{2}\ne P_{2}^{-1}}C_{P}x_{j}^{\dagger}H^{(P)}x_{i}^{(P)}-\sum_{P_{3}\ne P_{3}^{-1},P_{3}^{1}}C_{P}\lambda_{i}^{(P)}x_{j}^{\dagger}S^{(P)}x_{i}^{(P)})/(\lambda_{i}-\lambda_{j}).
\end{align}
Differentiate Eq. (\ref{eq:norm}), we have 
\[
\begin{split}
0  &=\sum_{P_{3}}C_{P}x_{i}^{(P)\dagger}S^{(P)}x_{i}^{(P)}\\
 &=\sum_{P_{3}\ne P_{3}^{-1},P_{3}^{1}}C_{P}x_{i}^{(P)\dagger}S^{(P)}x_{i}^{(P)}+x_{i}^{\dagger}Sx_{i}^{(P_{3}^{-1}(3))}+x_{i}^{(P_{3}^{1}(1))\dagger}Sx_{i}.
\end{split}
\]
Together with Eq. (\ref{eq:phase}), we have 
\[
x_{i}^{\dagger}Sx_{i}^{(P_{3}^{-1}(3))}=-\frac{1}{2}\sum_{P_{3}\ne P_{3}^{-1},P_{3}^{1}}C_{P}x_{i}^{(P)\dagger}S^{(P)}x_{i}^{(P)}.
\]
One can readily verify first order derivative in the main text is a
special case of these results.

\section*{References}
\bibliographystyle{iopart-num}
\bibliography{main}

\end{document}